\documentclass[a4paper,11pt]{article}
\usepackage{jheppub} 
\usepackage[T1]{fontenc}
\usepackage{multirow}
\usepackage{color}
\usepackage{ulem}
\usepackage{rotating}
\usepackage{booktabs}
\usepackage{dcolumn}
\usepackage{array} 
\usepackage{multirow}
\usepackage{lineno} 
\usepackage{siunitx}
\usepackage{amsmath}
\usepackage{bm}
\usepackage{mathrsfs}
\usepackage{indentfirst}

\DeclareSIUnit{\bmm}{\bm{m}}

\DeclareSIUnit{\clight}{\textnormal{\textit{c}}}

\newcolumntype{d}{D{.}{.}{-1}}
\newcolumntype{e}{D{.}{.}{8}}
\newcolumntype{f}{D{.}{.}{18}}
\newcolumntype{h}{D{.}{.}{13}}
\newcolumntype{g}{D{.}{.}{12}}

\title{\protect\boldmath Measurement of Born cross section of $e^+e^-\to\Sigma^+\bar\Sigma^-$ at center-of-mass energies between 3.510 and 4.951 GeV}
\collaborationImg{\includegraphics[width=0.25\textwidth]{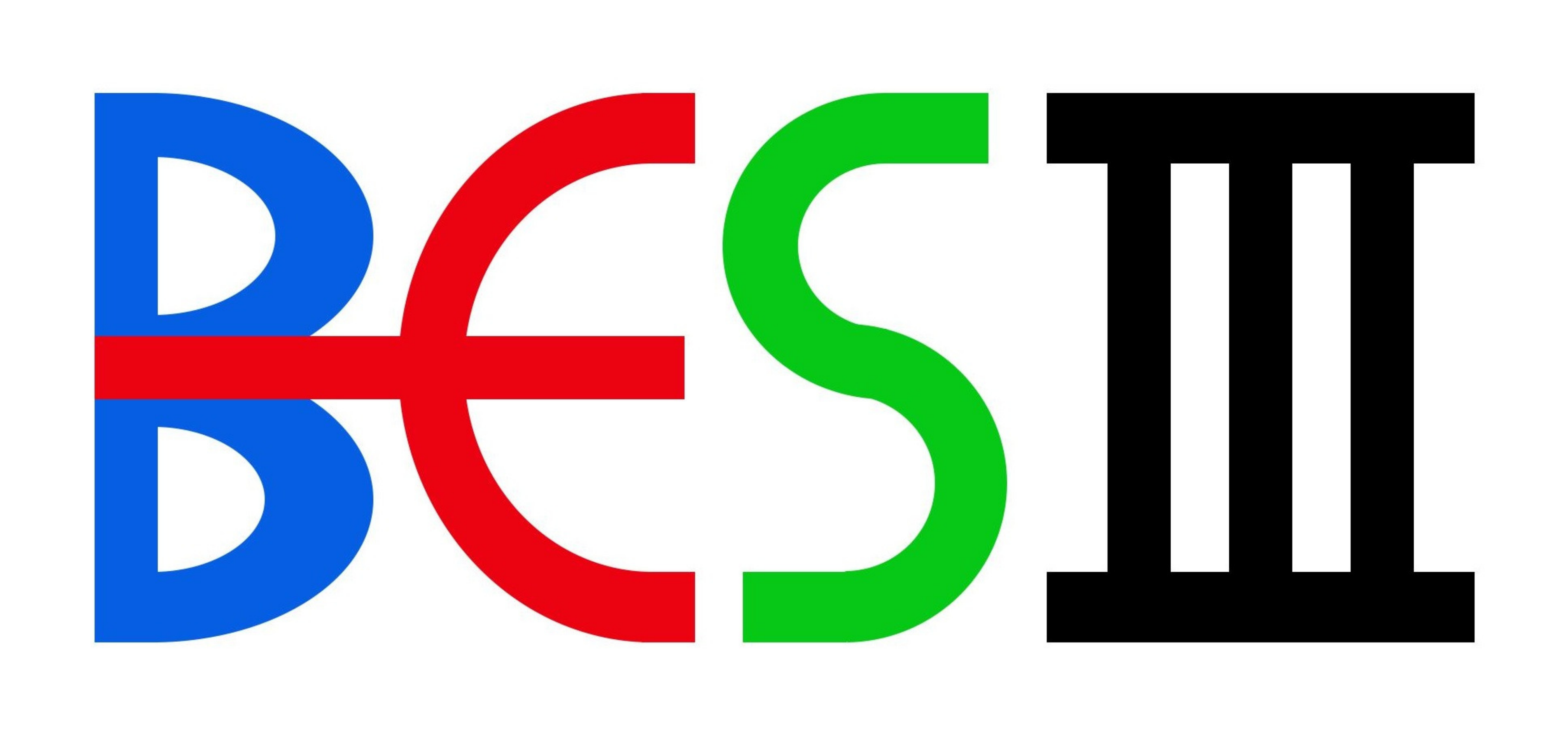}}
\collaboration{The BESIII collaboration}
\emailAdd{besiii-publications@ihep.ac.cn}

\begin{document} 
\abstract{Using 24.1 fb$^{-1}$ of $e^+e^-$ collision data collected with
  the BESIII detector at the BEPCII collider, the Born cross
  sections and effective form factors of the $e^+e^-\to\Sigma^+\bar\Sigma^-$ reaction are
  measured. The measurements are performed at center-of-mass energies
  ranging from 3.510 to 4.951 GeV. No significant evidence for the
  decay of the charmonium(-like)
  states, $\psi(3770)$, $\psi(4040)$, $\psi(4160)$, $Y(4230)$,
  $Y(4360)$, $\psi(4415)$, and $Y(4660)$, into a $\Sigma^+\bar\Sigma^-$ final state is
  observed. Consequently, upper limits for the products of the
  branching fractions and the electronic partial widths at the 90\%
  confidence level are reported for these decays. \\
 { K\textsc{eywords}: $e^{+}$-$e^{-}$ Experiment, QCD, Branching Fraction, Electroweak Interaction}
  }
\maketitle
\flushbottom    
\section{Introduction}
\label{sec:intro}
Below open-charm threshold, the mass spectrum of conventional
charmonium states aligns with the potential quark
model~\cite{Barnes:2005pb}. According to this model, there exist five
vector charmonium states between the 1D state ($\psi(3770)$) and 4.7
GeV/$c^2$, specifically identified as the 3S, 2D, 4S, 3D, and 5S
states~\cite{Farrar}. However, within this energy range, an overabundance 
of vector states have been detected. Among them, three
(conventional) states, namely $\psi(4040)$, $\psi(4160)$ and
$\psi(4415)$~\cite{BES:2001ckj}, the 3S, 2D, and 4S states,
respectively, are primarily characterized as open-charm states. 
In addition, four new (non-conventional) states, {\it i.e.} 
$Y(4230)$, $Y(4360)$, $Y(4634)$, and $Y(4660)$, are
predominantly observed in hidden-charm final states. These states are
generated via initial state radiation (ISR) processes at BaBar and
Belle~\cite{BaBar:2005hhc, Belle:2007dxy,
  BaBar:2006ait, Belle:2007umv, Belle:2008xmh, BaBar:2012hpr,
  Belle:2014wyt, BaBar:2012vyb, Belle:2013yex}, or by direct
production processes at CLEO~\cite{CLEO} and BESIII~\cite{BESIIIAB,
  BESIII:2023cmv}. The overpopulation of structures in this mass region and
the mismatch of the properties between the potential model predictions
and experimental measurements make them good candidates for exotic
states. Many hypotheses have been proposed to explain their
nature~\cite{Briceno,Farrar,Chen:2016qju,Wang:2019mhs, Close:2005iz,Qian:2021neg, Yan:2023yff, Dai:2023vsw, Wang:2023zxj},
including the possibility of being hybrid states, multiple-quark
states, or even molecular structures.  In particular, 
charmless decays of these non-conventional states are proposed by the hybrid model~\cite{Close:2005iz}.
On the other hand, if these states are considered as pure charmonium~\cite{Wang:2019mhs}, 
their baryonic decays, which have not yet been observed, would provide important information to validate the scenario as suggested in ref.~\cite{Qian:2021neg}. 

The complex situation reflects our limited understanding of the strong
interaction, particularly in its non-perturbative aspects. In order to
address this challenging problem, it is imperative to make additional
experimental measurements, and the study of $\psi/Y\to B\bar{B}$
decays holds great promise. These decays exhibit a straightforward
topology in terms of the final states, and the underlying interaction
mechanism is assumed to be dominated by three-gluon or one-photon
processes. Additionally, investigations into the electromagnetic form
factors or effective form factors of $B\bar{B}$ pairs have the
potential to provide insight into the internal composition of
charmonium(-like) states. Although many experimental
studies~\cite{Belle:2008xmh, Ablikim:2013pgf, BESIII:2021ccp, Ablikim:2019kkp,BESIII:2023rse, BESIII:2017kqg, Wang:2021lfq, Wang:2022bzl,Liu:2023xhg} of
$B\bar{B}$ pair production in this energy region have been performed by the
BESIII and Belle experiments, except for two evidences of
$\psi(3770)\to\Lambda\bar\Lambda$ and $\Xi^-\bar\Xi^+$, no significant indication for $B\bar{B}$ decay of
other vector charmonium(-like) states has been found. Thus, more
precise measurements of exclusive cross sections for $B\bar{B}$ final
states above the open-charm threshold are crucial.

This paper reports the measurements of the Born cross section and the
effective form factor for the process of $e^+e^-\to\Sigma^+\bar\Sigma^-$ using the data
corresponding to a total integrated luminosity of \SI{24.1}{fb^{-1}}
collected at center-of-mass (CM) energies ($\sqrt{s}$) between 3.510
and \SI{4.951}{GeV} with the BESIII detector~\cite{besiii} at the
BEPCII collider~\cite{BEPCII}. In addition, potential resonances
are searched for by fitting the dressed cross section of the $e^+e^-\to\Sigma^+\bar\Sigma^-$
reaction.

\section{BESIII Detector and Monte Carlo simulation}
The BESIII detector~\cite{besiii} records symmetric $e^+e^-$
collisions provided by the BEPCII storage ring~\cite{BEPCII} in the
range of $\sqrt{s}$ from 2.0 to \SI{4.95}{GeV}, with a peak luminosity
of \SI{1e33}{\per\centi\meter\squared\per\second} achieved at
$\sqrt{s} =$ \SI{3.77}{GeV}. BESIII has collected large data samples
in this energy region~\cite{Ablikim:2019hff, EcmsMea,
  EventFilter}. The cylindrical core of the BESIII detector covers
93\% of the full solid angle and consists of a helium-based multilayer
drift chamber~(MDC), a plastic scintillator time-of-flight
system~(TOF), and a CsI(Tl) electromagnetic calorimeter~(EMC), which
are all enclosed in a superconducting solenoidal magnet providing a
\SI{1.0}{T} magnetic field. The solenoid is supported by an octagonal
flux-return yoke with resistive plate counter muon identification
modules interleaved with steel. The charged-particle momentum
resolution at \SI{1}{GeV/\clight} is $0.5\%$, and the ${\rm d}E/{\rm
  d}x$ resolution is $6\%$ for electrons from Bhabha scattering. The
EMC measures photon energies with a resolution of $2.5\%$ ($5\%$) at
\SI{1}{GeV} in the barrel (end cap) region. The time resolution in the
TOF barrel region is \SI{68}{ps}, while that in the end cap region was
\SI{110}{ps}. The end cap TOF system was upgraded in 2015 using
multigap resistive plate chamber technology, providing a time
resolution of \SI{60}{ps}~\cite{etof1,etof2,etof3} and benefiting 82\%
of the data used in this analysis.

To evaluate detection efficiencies and estimate backgrounds, simulated
data samples are produced using {\sc geant4}-based Monte Carlo (MC)
software~\cite{GEANT4}, which incorporates the geometric description
of the BESIII detector~\cite{Huang:2022wuo} as well as the detector
response. The simulation of the  $e^+e^-\to\Sigma^+\bar\Sigma^-$ production process models
the beam energy spread in the $e^+e^-$
annihilation process, employing {\sc kkmc}~\cite{KKMC}. For each of
the 41 CM energy points ranging from 3.510 to 4.951 GeV, a sample of
100,000 events is simulated with a uniform phase space (PHSP)
distribution. The $\Sigma^+\bar\Sigma^-$ baryon pair and their subsequent decays are
simulated using {\sc
  evtgen}~\cite{evtgen2,EVTGEN} with a PHSP model.

\section{Event selection} Due to the large background in the selection
of  $e^+e^-\to\Sigma^+\bar\Sigma^-$ events, both the $\Sigma^+$ and
$\bar{\Sigma}^-$ are required to be reconstructed via the decay modes
$\Sigma^+\to p\pi^0$ and $\bar{\Sigma}^- \to \bar{p}\pi^0$ with the
subsequent decay $\pi^0\to\gamma\gamma$.

Tracks of charged particles detected in the MDC are required to
lie within the angular coverage of the MDC $\vert\!\cos\theta\vert < 0.93$, where
$\theta$ is the angle between the charged track and the $z$ axis,
which is the symmetry axis of the MDC. At least one positively charged
and one negatively charged track are required to be reconstructed
in the MDC with good Kalman fits. Because the proton and anti-proton can be
separated according to their momenta in a  $\Sigma^+\bar\Sigma^-$ decay, a charged
particle with momentum greater than 0.5 GeV/$c$ is identified as
a proton or anti-proton.

For $\pi^0$ reconstruction, the energies of photons are required to be
greater than 25 MeV in the EMC barrel region ($\vert\!\cos\theta\vert<0.8$) and
greater than 50 MeV in the EMC end cap ($0.86<\vert\!\cos\theta\vert<0.92$). To
suppress electronic noise and energy deposits unrelated to the events,
the EMC shower time measured with respect to the collision signal, is
required to satisfy $0 < t < 700$ ns. After these selections, at least
four photons are required.

\begin{figure}[!hbpt]
\begin{center}
\includegraphics[width=1.0\textwidth, trim=0 80 5 0, clip]{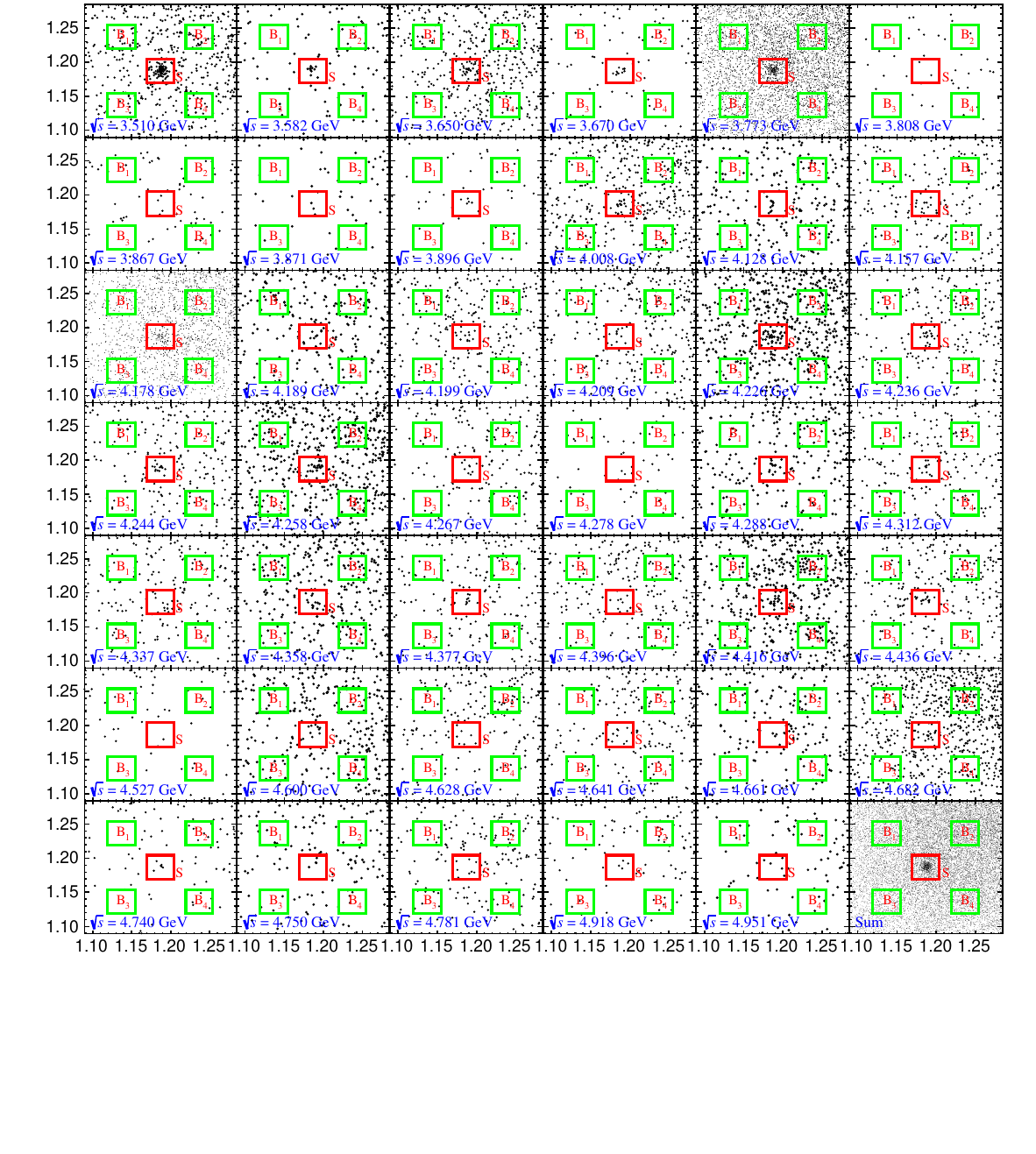}
\put(-230, 8){\boldmath $M_{p\pi^0}$ \textbf{(GeV/$c^2$)}}
\put(-430, 185){\rotatebox{90}{\boldmath $M_{\bar{p}\pi^0}$ \textbf{(GeV/$c^2$)}}}
\end{center}
\caption{ Distributions of $M_{p\pi^0}$ versus $M_{\bar{p}\pi^0}$ for
  data at each energy point between 3.510 and \SI{4.951}{GeV} and the
  sum of all energy points (bottom right) from data. The red boxes
  represent the signal regions and the green boxes represent the
  selected sideband regions.}
\label{Fig:SUM:DATA}
\end{figure}

The best candidate of all combinations of
$p\bar{p}\gamma\gamma\gamma\gamma$ within an event is determined by a
six-constraint (6C) kinematic fit, which imposes energy and momentum
conservation and constrains the masses of photon pairs to the known mass of
$\pi^0$~\cite{PDG2020}. The $p\bar{p}\pi^0\pi^0$ combination with the smallest
fit $\chi^2$ is chosen. For different $p(\bar{p})$ and $\pi^0$
combinations, the $\Sigma^+$ and $\bar{\Sigma}^-$ pair with the
minimum of
$\sqrt{(M_{p\pi^0}-m_{\Sigma^+})^2+(M_{\bar{p}\pi^0}-m_{\bar{\Sigma}^-})^2}$,
is selected. Here, $M_{p\pi^0(\bar{p}\pi^0)}$ is the invariant mass of
the $p\pi^0(\bar{p}\pi^0)$ combination, and
$m_{\Sigma^+(\bar{\Sigma}^-)}$ is the known mass of
$\Sigma^+(\bar{\Sigma}^-)$ from the Particle Data Group
(PDG)~\cite{PDG2020}. Figure~\ref{Fig:SUM:DATA} shows the
distributions of $M_{p\pi^0}$ versus $M_{\bar{p}\pi^0}$ for each
energy point and the sum of all energy
points. $M_{p\pi^0(\bar{p}\pi^0)}$ is required to be within the range
of $[m_{\Sigma^+} - 4\sigma, m_{\Sigma^+} + 3\sigma]$, which is labeled
by $S$ in Fig.~\ref{Fig:SUM:DATA}. The resolution $\sigma$ and the
signal region are determined by a fit with the Crystal-Ball
function~\cite{Oreglia:1980cs}. Due to the longer tail of the photon
energy deposition at the low energy side in the EMC, the signal region
is asymmetric.

\section{Born cross section measurement}
\subsection{Determination of signal yields}

After applying the event selection criteria on data, the remaining
background mainly comes from non-$\Sigma^+\bar{\Sigma}^-$ events, such
as $e^+e^-\to\pi^{0}\pi^{0}J/\psi\to\pi^0\pi^0p\bar{p}$. To estimate
the background yield in the signal region, four sideband regions $B_i$
(where $i =$ 1, 2, 3, 4) are utilized. These sideband regions, shown
in figure~\ref{Fig:SUM:DATA}, have the same area as the signal region,
and the exact ranges are defined by
\begin{itemize}
 \item $B_{1}$: $M_{p\pi^0}\in$ [1.119, 1.154] GeV/$c^2$ $\&$ $M_{\bar{p}\pi^0}\in$ [1.219, 1.254] GeV/$c^2$, 
 \item $B_{2}$: $M_{p\pi^0}\in$ [1.219, 1.254] GeV/$c^2$ $\&$ $M_{\bar{p}\pi^0}\in$ [1.219, 1.254] GeV/$c^2$, 
 \item $B_{3}$: $M_{p\pi^0}\in$ [1.119, 1.154] GeV/$c^2$ $\&$ $M_{\bar{p}\pi^0}\in$ [1.119, 1.154] GeV/$c^2$, 
 \item $B_{4}$: $M_{p\pi^0}\in$ [1.219, 1.254] GeV/$c^2$ $\&$ $M_{\bar{p}\pi^0}\in$ [1.119, 1.154] GeV/$c^2$.
\end{itemize}
The signal yield $N_{\rm obs}$ for the  $e^+e^-\to\Sigma^+\bar\Sigma^-$ reaction at each
energy point is determined by subtracting the number of
events in the sideband regions from the signal region, {\it i.e.}, $N_{\rm
  obs} = N_{S} - N_{\rm bkg}$, where $N_{S}$ is the number of events
from the signal region and $N_{\rm
  bkg}=\frac{1}{4}\sum^{4}_{i=1}N_{B_{i}}$. Statistical
uncertainties are calculated based on the TRolke
method~\cite{Lundberg:2009iu}, and the statistical significance is
evaluated based on the observed $p$-value~\cite{p:value}. The results
are listed in table~\ref{tab:signal:yields:DD01}. For the energy
points with statistical significance less than $3\sigma$, the
upper limit at the 90\% confidence level (C.L.) is also estimated
based on the TRolke method, which takes into account systematic
uncertainties.

\begin{table}[!hpt]
	\caption{\small Number of events: $N_S$ is from the signal
          region, $N_{\rm bkg}$ is the number of background events,
          $N_{\rm obs}$ is the number of events by subtracting the
          backgrounds, $S(\sigma)$ is the statistical significance,
          and $N^{\rm UL}$ is the upper limit for an energy point
          with statistical significance less than $3\sigma$.}
        \centering \scalebox{0.95}{
        \begin{tabular}{lllll}
        \hline
        \hline
        \multicolumn{1}{c}{$\sqrt{s}$ (GeV)} & \multicolumn{1}{c}{$N_{S}$} & \multicolumn{1}{l}{$N_{\rm bkg}$} & \multicolumn{1}{l}{$N_{\rm obs}$ ($\textless N^{\rm UL}$)} & \multicolumn{1}{c}{$\mathcal{S}$ $(\sigma)$} \\ 
        \hline
        3.510    & $89.0_{-9.4}^{+10.5}    $ & $18.8_{-4.6}^{+5.1}     $ & $70.3 _{-8.4 }^{+10.5}                    $   & 7.9   \\
        3.582    & $12.0_{-3.4}^{+4.6}     $ & $4.0_{-1.9}^{+3.2}      $ & $8.0  _{-3.1 }^{+3.8}                     $   & 3.3   \\
        3.650    & $44.0_{-6.6}^{+7.7}     $ & $10.0_{-3.1}^{+4.3}     $ & $34.0 _{-6.3 }^{+7.0}                     $   & 7.9   \\
        3.670    & $13.0_{-3.6}^{+4.7}     $ & $2.8_{-1.8}^{+2.5}      $ & $10.3 _{-2.5 }^{+4.7}                     $   & 4.5   \\
        3.773    & $324.0_{-18.0}^{+19.0}  $ & $139.3_{-12.1}^{+12.5}  $ & $184.8_{-17.4}^{+18.6}                    $   & 7.9   \\
        3.808    & $2.0_{-1.3}^{+2.6}      $ & $1.3_{-1.1}^{+2.0}      $ & $0.8  _{-0.8 }^{+2.0}   ~(\textless 4.5 ) $   & 0.9   \\
        3.867    & $4.0_{-1.9}^{+3.2}      $ & $3.3_{-1.9}^{+2.6}      $ & $0.8  _{-0.8 }^{+2.6}   ~(\textless 5.8 ) $   & 0.8   \\
        3.871    & $4.0_{-1.9}^{+3.2}      $ & $1.0_{-0.8}^{+2.3}      $ & $3.0  _{-1.7 }^{+2.3}   ~(\textless 7.5 ) $   & 2.3   \\
        3.896    & $6.0_{-2.4}^{+3.6}      $ & $1.5_{-1.1}^{+2.5}      $ & $4.5  _{-1.6 }^{+3.3}   ~(\textless 10.2) $   & 2.8   \\
        4.008    & $29.0_{-5.4}^{+6.4}     $ & $11.3_{-3.6}^{+4.1}     $ & $17.8 _{-4.8 }^{+6.0}                     $   & 4.5   \\
        4.128    & $13.0_{-3.6}^{+4.7}     $ & $6.5_{-2.5}^{+3.7}      $ & $6.5  _{-2.8 }^{+4.4}   ~(\textless 14.8) $   & 2.4   \\
        4.157    & $16.0_{-4.0}^{+5.1}     $ & $4.3_{-2.2}^{+2.9}      $ & $11.8 _{-3.4 }^{+4.6}                     $   & 4.4   \\
        4.178    & $99.0_{-9.9}^{+11.0}    $ & $54.5_{-7.4}^{+8.4}     $ & $44.5 _{-9.1 }^{+10.8}                    $   & 5.5   \\
        4.189    & $12.0_{-3.4}^{+4.6}     $ & $9.0_{-2.9}^{+4.1}      $ & $3.0  _{-3.0 }^{+3.8}   ~(\textless 10.9) $   & 1.3   \\
        4.199    & $15.0_{-3.8}^{+5.0}     $ & $7.8_{-3.0}^{+3.6}      $ & $7.3  _{-2.8 }^{+5.0}   ~(\textless 16.3) $   & 2.5   \\
        4.209    & $14.0_{-3.7}^{+4.8}     $ & $7.8_{-3.0}^{+3.6}      $ & $6.3  _{-2.7 }^{+4.8}   ~(\textless 15.1) $   & 2.2   \\
        4.226    & $45.0_{-6.7}^{+7.8}     $ & $16.0_{-4.0}^{+5.1}     $ & $29.0 _{-6.4 }^{+7.0}                     $   & 6.0   \\
        4.236    & $21.0_{-4.5}^{+5.7}     $ & $9.0_{-2.9}^{+4.1}      $ & $12.0 _{-4.3 }^{+4.9}                     $   & 3.5   \\
        4.244    & $16.0_{-4.0}^{+5.1}     $ & $6.8_{-2.8}^{+3.4}      $ & $9.3  _{-2.9 }^{+5.1}                     $   & 3.1   \\
        4.258    & $28.0_{-5.3}^{+6.4}     $ & $19.8_{-4.7}^{+5.2}     $ & $8.3  _{-4.2 }^{+6.4}   ~(\textless 20.7) $   & 2.0   \\
        4.267    & $13.0_{-3.6}^{+4.7}     $ & $7.0_{-2.6}^{+3.8}      $ & $6.0  _{-3.3 }^{+3.9}   ~(\textless 13.9) $   & 2.2   \\
        4.278    & $5.0_{-2.2}^{+3.4}      $ & $3.0_{-1.6}^{+2.9}      $ & $2.0  _{-2.0 }^{+2.6}   ~(\textless 7.2 ) $   & 1.3   \\
        4.288    & $17.0_{-4.1}^{+5.2}     $ & $8.3_{-3.1}^{+3.6}      $ & $8.8  _{-3.5 }^{+4.7}   ~(\textless 17.9) $   & 2.8   \\
        4.312    & $15.0_{-3.8}^{+5.0}     $ & $8.0_{-2.8}^{+3.9}      $ & $7.0  _{-3.5 }^{+4.2}   ~(\textless 15.5) $   & 2.4   \\
        4.337    & $13.0_{-3.6}^{+4.7}     $ & $5.8_{-2.6}^{+3.2}      $ & $7.3  _{-2.5 }^{+4.7}   ~(\textless 15.7) $   & 2.7   \\
        4.358    & $16.0_{-4.0}^{+5.1}     $ & $13.8_{-3.9}^{+4.5}     $ & $2.3  _{-2.2 }^{+5.1}   ~(\textless 12.2) $   & 1.0   \\
        4.377    & $14.0_{-3.7}^{+4.8}     $ & $6.3_{-2.7}^{+3.3}      $ & $7.8  _{-3.2 }^{+4.3}   ~(\textless 16.0) $   & 2.8   \\
        4.396    & $11.0_{-3.3}^{+4.4}     $ & $7.3_{-2.9}^{+3.5}      $ & $3.8  _{-3.8 }^{+3.9}   ~(\textless 11.5) $   & 1.6   \\
        4.416    & $26.0_{-5.1}^{+6.2}     $ & $19.0_{-4.3}^{+5.4}     $ & $7.0  _{-4.8 }^{+5.4}   ~(\textless 18.4) $   & 1.8   \\
        4.436    & $17.0_{-4.1}^{+5.2}     $ & $8.8_{-3.2}^{+3.7}      $ & $8.3  _{-3.0 }^{+5.2}   ~(\textless 17.9) $   & 2.6   \\
        4.527    & $3.0_{-1.6}^{+2.9}      $ & $2.5_{-1.5}^{+2.8}      $ & $0.5  _{-0.5 }^{+2.6}   ~(\textless 5.2 ) $   & 0.7   \\
        4.600    & $11.0_{-3.3}^{+4.4}     $ & $10.5_{-3.2}^{+4.3}     $ & $0.5  _{-0.5 }^{+4.2}   ~(\textless 8.8 ) $   & 0.7   \\
        4.628    & $7.0_{-2.6}^{+3.8}      $ & $6.5_{-2.5}^{+3.7}      $ & $0.5  _{-0.5 }^{+3.5}   ~(\textless 7.3 ) $   & 0.7   \\
        4.641    & $12.0_{-3.4}^{+4.6}     $ & $6.3_{-2.7}^{+3.3}      $ & $5.8  _{-2.9 }^{+4.1}   ~(\textless 13.6) $   & 2.2   \\
        4.661    & $8.0_{-2.8}^{+3.9}      $ & $6.0_{-2.4}^{+3.6}      $ & $2.0  _{-2.0 }^{+3.2}   ~(\textless 8.6 ) $   & 1.1   \\
        4.682    & $26.0_{-5.1}^{+6.2}     $ & $21.8_{-4.9}^{+5.4}     $ & $4.3  _{-4.2 }^{+6.2}   ~(\textless 16.6) $   & 1.3   \\
        4.740    & $7.0_{-2.6}^{+3.8}      $ & $3.3_{-1.9}^{+2.6}      $ & $3.8  _{-2.1 }^{+3.2}   ~(\textless 9.9 ) $   & 2.0   \\
        4.750    & $4.0_{-1.9}^{+3.2}      $ & $3.5_{-1.8}^{+3.0}      $ & $0.5  _{-0.5 }^{+2.8}   ~(\textless 5.8 ) $   & 0.7   \\
        4.781    & $10.0_{-3.1}^{+4.3}     $ & $6.5_{-2.5}^{+3.7}      $ & $3.5  _{-2.3 }^{+4.0}   ~(\textless 11.1) $   & 1.5   \\
        4.918    & $5.0_{-2.2}^{+3.4}      $ & $4.5_{-2.0}^{+3.3}      $ & $0.5  _{-0.5 }^{+3.1}   ~(\textless 6.4 ) $   & 0.7   \\
        4.951    & $5.0_{-2.2}^{+3.4}      $ & $2.0_{-1.3}^{+2.6}      $ & $3.0  _{-1.9 }^{+2.6}   ~(\textless 8.0 ) $   & 1.9  \\
        \hline
        \hline
        \end{tabular}}
        \label{tab:signal:yields:DD01}
\end{table}

\subsection{Determination of Born cross section and effective form
factor} 
\begin{figure}[hbpt]
	\begin{center}
	\includegraphics[width=0.75\textwidth]{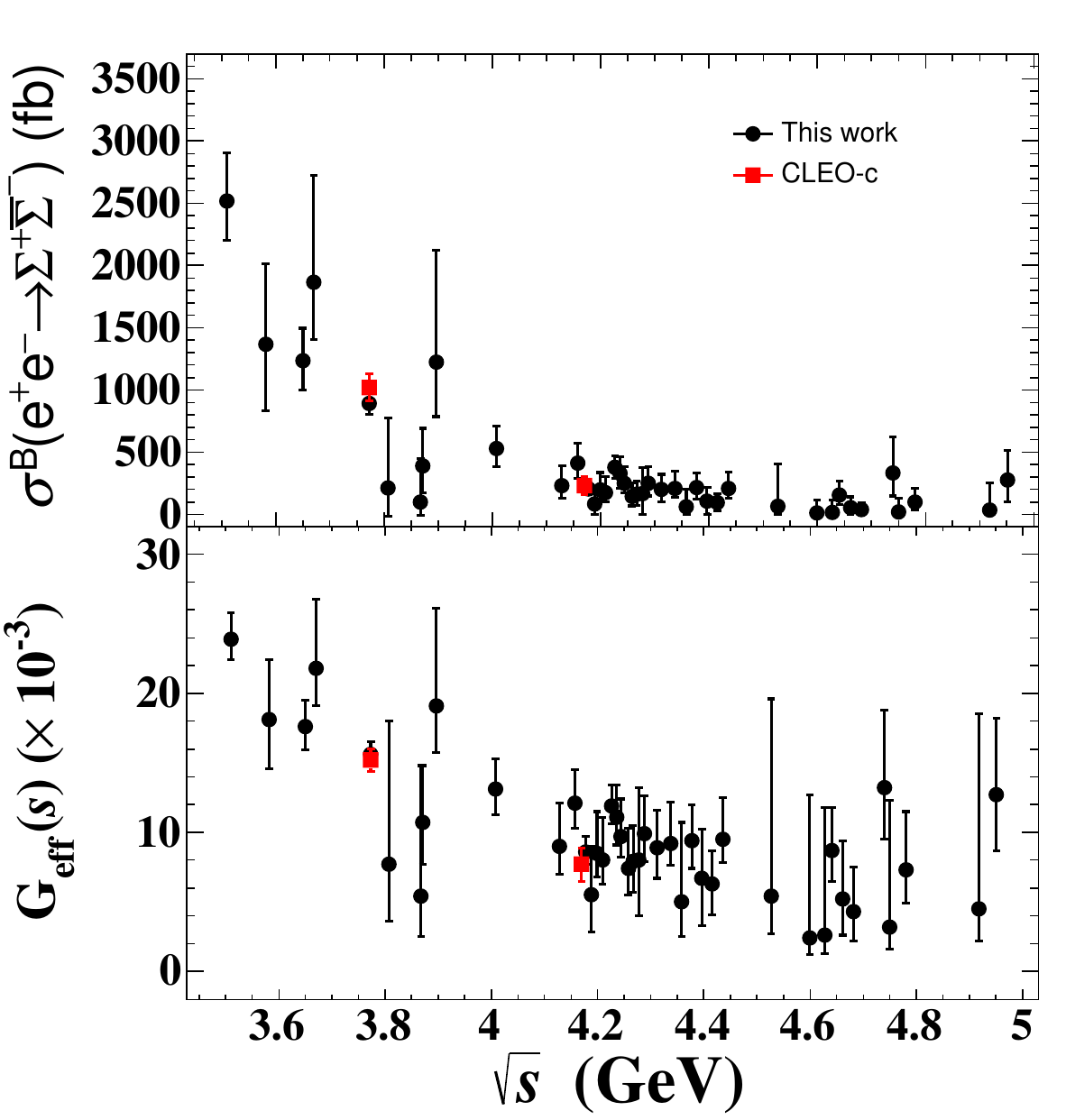}
	\end{center}
	\caption{The measured Born cross section (top) and $\Sigma^+$
          effective form factor (bottom) for  $e^+e^-\to\Sigma^+\bar\Sigma^-$ versus CM
          energy, where the uncertainties include both the statistical
          and systematic ones. }
	\label{Fig:XiXi::CS:BCS_VS_EFF}
\end{figure}

The Born cross section for the  $e^+e^-\to\Sigma^+\bar\Sigma^-$ process at a given
CM energy is calculated by \begin{equation} \sigma^{B} =\frac{N_{\rm
obs}}{{\cal{L}}\cdot(1 + \delta)\cdot\frac{1}{|1 -
\prod|^{2}}\cdot\epsilon\cdot {\cal B}^2_{\Sigma^+\to p\pi^0} \cdot
{\cal B}^2_{\pi^0\to\gamma\gamma}}, \end{equation} where $N_{\rm obs}$
is the number of observed signal events, ${\cal{L}}$ is the integrated
luminosity, $(1 + \delta)$ is the ISR correction factor,
$\frac{1}{|1-\Pi|^2}$ is the vacuum polarization (VP) correction
factor, $\epsilon$ is the detection efficiency, and ${\cal
B}_{\Sigma^+\to p\pi^0}$ and ${\cal B}_{\pi^0\to\gamma\gamma}$ are the
PDG branching fractions~\cite{PDG2020}. The ISR correction factor is
obtained using the QED calculation as described in
ref.~\cite{Kuraev:1985hb}, and the VP correction factor is calculated
according to ref.~\cite{Jegerlehner:2011ti}. The efficiency and ISR
correction factor are obtained through an iterative
process. Initially, the cross section is measured without any
correction factors. Using this initial line shape, signal MC samples
are regenerated, and their efficiencies and ISR correction factors are
recalculated. Subsequently, the Born cross section of  $e^+e^-\to\Sigma^+\bar\Sigma^-$ is
updated and used as input for the next iteration. To expedite the
iteration procedure, an iterative weighting method, as proposed in
ref.~\cite{Sun:2020ehv}, is employed. The procedure is iteratively
performed until the difference of $\epsilon\cdot(1+\delta)$ with the
last iteration falls below 0.5\%. The results of the measured Born
cross sections and the $\Sigma^+$ effective form factors $G_{\rm
eff}(s)$ for different energy points are listed in
table~\ref{tab:signal:yields:DD}. $G_{\rm eff}(s)$ is defined
as~\cite{Ablikim:2019kkp} \begin{equation}
        |G_{\rm eff}(s)| =
        \sqrt{\frac{3s\tau\sigma^B}{2\pi\alpha^2C\beta(2\tau+1)}},
\end{equation}
where $s$ is the square of the CM energy,
$\alpha = \frac{1}{137}$ is the fine structure constant, the variable
$\beta = \sqrt{1-\frac{1}{\tau}}$ is the velocity of 
$\Sigma^+$ in the laboratory department, $\tau =
\frac{s}{4m_{\Sigma^+}^2}$, 
and the Coulomb factor
$C$~\cite{Baldini, Arbuzov} parameterizes the electromagnetic
interaction between the outgoing baryon and anti-baryon. For neutral
baryons, the Coulomb factor is unity, while for point-like charged
fermions,
$C=\frac{\pi\alpha}{\beta}\cdot\frac{\sqrt{1-\beta^2}}{1-e^{-\frac{\pi\alpha}{\beta}}}$
\cite{Sommerfeld, Sakharov,Tzara, Wang:2022zyc}.
Figure~\ref{Fig:XiXi::CS:BCS_VS_EFF} displays the energy dependence of
the Born cross section and $\Sigma^+$ effective form factor $G_{\rm
  eff}(s)$ measurements, as well as a comparison of the Born cross
sections and the effective form factors with the CLEO-c
results~\cite{Dobbs:2017hyd} at $\sqrt{s} = 3.770$ and $4.160$ GeV.

\begin{table}[!hpt]
	\begin{center}
	\caption{\small The CM energy ($\sqrt{s}$), the integrated luminosity (${\cal{L}}$), the VP correction factor ($\frac{1}{|1 - \prod|^{2}}$), the ISR correction factor and the detection efficiency ($\epsilon\cdot(1 + \delta)$), the signal yield ($N_{\rm obs}$), the upper limit of signal yield at the 90\% C.L. ($N^{\rm UL}$), the Born cross section ($\sigma^{B}$), the effective from factor ($|G_{\rm eff}(s)|$) and the statistical significance ($\mathcal{S}$). The first and second uncertainties for $\sigma^{B}$ and $|G_{\rm eff}(s)|$ are statistical and systematic, respectively.}
    \scalebox{0.71}{
        \begin{tabular}{l@{\hspace{-0.5cm}}dccgf@{\hspace{0.5cm}}h@{\hspace{0.5cm}}d} 
        \hline 
        \hline
        \multicolumn{1}{c}{$\sqrt{s}$ (GeV)} &\multicolumn{1}{c}{${\cal{L}}$ (pb$^{-1})$} &\multicolumn{1}{c}{\,\,$\frac{1}{|1 - \prod|^{2}}$} &\multicolumn{1}{c}{$\epsilon\cdot(1+\delta)$} &\multicolumn{1}{c}{$N_{\rm obs}$ ($\textless N^{\rm UL}$)} &\multicolumn{1}{c}{$\sigma^{B}$ (fb)} &\multicolumn{1}{c}{$|G_{\rm eff}(s)|$ $\times 10^{-3}$} &\multicolumn{1}{c}{$\mathcal{S}$ $(\sigma)$} \\ 
        \hline
        3.510 & 405.4  & 1.04 & 0.25 &70.3 _{-8.4 }^{+10.5}                    &2517 .9^{+376.3}_{-301.1} \pm103.2                      &23 .9^{+1.8 }_{-1.4}  \pm0.5                      &~7.9\\  
        3.582 & 85.7   & 1.04 & 0.25 &8.0  _{-3.1 }^{+3.8}                     &1366 .3^{+649.0}_{-529.4} \pm56.0                       &18 .1^{+4.3 }_{-3.5}  \pm0.4                      &~3.3\\
        3.650 & 410.0  & 1.02 & 0.25 &34.0 _{-6.3 }^{+7.0}                     &1236 .5^{+254.6}_{-229.1} \pm50.7                       &17 .6^{+1.8 }_{-1.6}  \pm0.4                      &~7.9\\
        3.670 & 84.7   & 0.99 & 0.25 &10.3 _{-2.5 }^{+4.7}                     &1866 .2^{+855.7}_{-455.2} \pm76.5                       &21 .8^{+5.0 }_{-2.7}  \pm0.4                      &~4.5\\
        3.773 & 2931.8 & 1.06 & 0.26 &184.8_{-17.4}^{+18.6}                    &891  .4^{+89.7} _{-84.0}  \pm36.5                       &15 .6^{+0.8 }_{-0.7}  \pm0.3                      &~7.9\\
        3.808 & 50.5   & 1.06 & 0.26 &0.8  _{-0.8 }^{+2.0}   ~(\textless 4.5 ) &211  .4^{+563.8}_{-225.5} \pm8.7   ~(\textless 1268.5)  &7  .7^{+10.3}_{-4.1}  \pm0.2   ~(\textless 18.9)  &~0.9\\
        3.867 & 108.9  & 1.05 & 0.25 &0.8  _{-0.8 }^{+2.6}   ~(\textless 5.8 ) &99   .4^{+344.7}_{-106.1} \pm4.1   ~(\textless 768.9 )  &5  .4^{+9.4 }_{-2.9}  \pm0.1   ~(\textless 15.0)  &~0.8\\
        3.871 & 110.3  & 1.05 & 0.26 &3.0  _{-1.7 }^{+2.3}   ~(\textless 7.5 ) &390  .4^{+299.3}_{-221.2} \pm16.0  ~(\textless 976.1 )  &10 .7^{+4.1 }_{-3.0}  \pm0.2   ~(\textless 17.0)  &~2.3\\
        3.896 & 52.6   & 1.05 & 0.26 &4.5  _{-1.6 }^{+3.3}   ~(\textless 10.2) &1222 .3^{+896.4}_{-434.6} \pm50.1  ~(\textless 2770.5)  &19 .1^{+7.0 }_{-3.4}  \pm0.4   ~(\textless 28.8)  &~2.8\\
        4.008 & 482.0  & 1.04 & 0.26 &17.8 _{-4.8 }^{+6.0}                     &527  .6^{+178.4}_{-142.7} \pm21.6                       &13 .1^{+2.2 }_{-1.8}  \pm0.3                      &~4.5\\
        4.128 & 401.5  & 1.05 & 0.26 &6.5  _{-2.8 }^{+4.4}   ~(\textless 14.8) &231  .3^{+156.6}_{-99.6}  \pm9.5   ~(\textless 526.6 )  &9  .0^{+3.1 }_{-1.9}  \pm0.2   ~(\textless 13.6)  &~2.4\\
        4.157 & 408.7  & 1.05 & 0.26 &11.8 _{-3.4 }^{+4.6}                     &411  .5^{+161.1}_{-119.1} \pm16.9                       &12 .1^{+2.4 }_{-1.8}  \pm0.2                      &~4.4\\
        4.178 & 3189.0 & 1.05 & 0.25 &44.5 _{-9.1 }^{+10.8}                    &202  .7^{+49.2} _{-41.5}  \pm8.3                        &8  .6^{+1.0 }_{-0.9}  \pm0.2                      &~5.5\\
        4.189 & 526.7  & 1.06 & 0.25 &3.0  _{-3.0 }^{+3.8}   ~(\textless 10.9) &82   .3^{+104.2}_{-82.3}  \pm3.4   ~(\textless 298.9 )  &5  .5^{+3.5 }_{-2.7}  \pm0.1   ~(\textless 10.5)  &~1.3\\
        4.199 & 526.0  & 1.06 & 0.25 &7.3  _{-2.8 }^{+5.0}   ~(\textless 16.3) &198  .0^{+136.6}_{-76.5}  \pm8.1   ~(\textless 445.2 )  &8  .5^{+2.9 }_{-1.6}  \pm0.2   ~(\textless 12.8)  &~2.5\\
        4.209 & 517.1  & 1.06 & 0.25 &6.3  _{-2.7 }^{+4.8}   ~(\textless 15.1) &173  .1^{+133.0}_{-74.8}  \pm7.1   ~(\textless 418.3 )  &8  .0^{+3.1 }_{-1.7}  \pm0.2   ~(\textless 12.5)  &~2.2\\
        4.226 & 1100.9 & 1.06 & 0.25 &29.0 _{-6.4 }^{+7.0}                     &377  .3^{+91.1} _{-83.3}  \pm15.5                       &11 .9^{+1.4 }_{-1.3}  \pm0.2                      &~6.0\\
        4.236 & 530.3  & 1.06 & 0.25 &12.0 _{-4.3 }^{+4.9}                     &327  .4^{+133.7}_{-117.3} \pm13.4                       &11 .1^{+2.3 }_{-2.0}  \pm0.2                      &~3.5\\
        4.244 & 538.1  & 1.06 & 0.25 &9.3  _{-2.9 }^{+5.1}                     &248  .3^{+136.9}_{-77.9}  \pm10.2                       &9  .7^{+2.7 }_{-1.5}  \pm0.2                      &~3.1\\
        4.258 & 828.4  & 1.05 & 0.25 &8.3  _{-4.2 }^{+6.4}   ~(\textless 20.7) &143  .0^{+111.0}_{-72.8}  \pm5.9   ~(\textless 358.9 )  &7  .4^{+2.9 }_{-1.9}  \pm0.2   ~(\textless 11.7)  &~2.0\\
        4.267 & 531.1  & 1.05 & 0.25 &6.0  _{-3.3 }^{+3.9}   ~(\textless 13.9) &162  .7^{+105.8}_{-89.5}  \pm6.7   ~(\textless 377.0 )  &7  .9^{+2.6 }_{-2.2}  \pm0.2   ~(\textless 12.1)  &~2.2\\
        4.278 & 175.7  & 1.05 & 0.25 &2.0  _{-2.0 }^{+2.6}   ~(\textless 7.2 ) &164  .9^{+214.3}_{-164.9} \pm6.8   ~(\textless 593.6 )  &8  .0^{+5.2 }_{-4.0}  \pm0.2   ~(\textless 15.2)  &~1.3\\
        4.288 & 502.4  & 1.05 & 0.26 &8.8  _{-3.5 }^{+4.7}   ~(\textless 17.9) &249  .3^{+133.9}_{-99.7}  \pm10.2  ~(\textless 509.9 )  &9  .9^{+2.7 }_{-2.0}  \pm0.2   ~(\textless 14.1)  &~2.8\\
        4.312 & 501.2  & 1.05 & 0.25 &7.0  _{-3.5 }^{+4.2}   ~(\textless 15.5) &201  .4^{+120.8}_{-100.7} \pm8.3   ~(\textless 445.9 )  &8  .9^{+2.7 }_{-2.2}  \pm0.2   ~(\textless 13.3)  &~2.4\\
        4.337 & 505.0  & 1.05 & 0.25 &7.3  _{-2.5 }^{+4.7}   ~(\textless 15.7) &208  .6^{+135.2}_{-71.9}  \pm8.6   ~(\textless 451.7 )  &9  .2^{+3.0 }_{-1.6}  \pm0.2   ~(\textless 13.5)  &~2.7\\
        4.358 & 544.0  & 1.05 & 0.24 &2.3  _{-2.2 }^{+5.1}   ~(\textless 12.2) &62   .0^{+140.5}_{-60.6}  \pm2.5   ~(\textless 336.1 )  &5  .0^{+5.7 }_{-2.5}  \pm0.1   ~(\textless 11.7)  &~1.0\\
        4.377 & 522.7  & 1.05 & 0.25 &7.8  _{-3.2 }^{+4.3}   ~(\textless 16.0) &214  .7^{+119.1}_{-88.7}  \pm8.8   ~(\textless 443.3 )  &9  .4^{+2.6 }_{-1.9}  \pm0.2   ~(\textless 13.6)  &~2.8\\
        4.396 & 507.8  & 1.05 & 0.25 &3.8  _{-3.8 }^{+3.9}   ~(\textless 11.5) &107  .8^{+112.1}_{-109.3} \pm4.4   ~(\textless 330.7 )  &6  .7^{+3.5 }_{-3.4}  \pm0.1   ~(\textless 11.8)  &~1.6\\
        4.416 & 1090.7 & 1.05 & 0.25 &7.0  _{-4.8 }^{+5.4}   ~(\textless 18.4) &93   .6^{+72.2} _{-64.2}  \pm3.8   ~(\textless 245.9 )  &6  .3^{+2.4 }_{-2.2}  \pm0.1   ~(\textless 10.2)  &~1.8\\
        4.436 & 569.9  & 1.05 & 0.25 &8.3  _{-3.0 }^{+5.2}   ~(\textless 17.9) &208  .6^{+131.5}_{-75.9}  \pm8.6   ~(\textless 452.7 )  &9  .5^{+3.0 }_{-1.7}  \pm0.2   ~(\textless 14.0)  &~2.6\\
        4.527 & 112.1  & 1.05 & 0.25 &0.5  _{-0.5 }^{+2.6}   ~(\textless 5.2 ) &65   .0^{+337.9}_{-65.0}  \pm2.7   ~(\textless 675.7 )  &5  .4^{+14.2}_{-2.7}  \pm0.1   ~(\textless 17.6)  &~0.7\\
        4.600 & 586.9  & 1.05 & 0.25 &0.5  _{-0.5 }^{+4.2}   ~(\textless 8.8 ) &12   .5^{+105.3}_{-12.5}  \pm0.5   ~(\textless 220.6 )  &2  .4^{+10.3}_{-1.2}  \pm0.1   ~(\textless 10.3)  &~0.7\\
        4.628 & 521.5  & 1.05 & 0.25 &0.5  _{-0.5 }^{+3.5}   ~(\textless 7.3 ) &14   .2^{+99.2} _{-14.2}  \pm0.6   ~(\textless 206.9 )  &2  .6^{+9.2 }_{-1.3}  \pm0.1   ~(\textless 10.0)  &~0.7\\
        4.641 & 551.7  & 1.05 & 0.25 &5.8  _{-2.9 }^{+4.1}   ~(\textless 13.6) &154  .3^{+110.0}_{-77.8}  \pm6.2   ~(\textless 364.9 )  &8  .7^{+3.1 }_{-2.2}  \pm0.2   ~(\textless 13.4)  &~2.2\\
        4.661 & 529.4  & 1.05 & 0.25 &2.0  _{-2.0 }^{+3.2}   ~(\textless 8.6 ) &54   .5^{+87.2} _{-54.5}  \pm2.2   ~(\textless 234.3 )  &5  .2^{+4.2 }_{-2.6}  \pm0.1   ~(\textless 10.8)  &~1.1\\
        4.682 & 1667.4 & 1.05 & 0.25 &4.3  _{-4.2 }^{+6.2}   ~(\textless 16.6) &37   .4^{+54.6} _{-37.0}  \pm1.5   ~(\textless 146.3 )  &4  .3^{+3.2 }_{-2.1}  \pm0.1   ~(\textless 8.6 )  &~1.3\\
        4.740 & 163.9  & 1.05 & 0.25 &3.8  _{-2.1 }^{+3.2}   ~(\textless 9.9 ) &333  .9^{+284.9}_{-187.0} \pm13.4  ~(\textless 881.4 )  &13 .2^{+5.6 }_{-3.7}  \pm0.3   ~(\textless 21.4)  &~2.0\\
        4.750 & 366.6  & 1.05 & 0.25 &0.5  _{-0.5 }^{+2.8}   ~(\textless 5.8 ) &19   .9^{+111.6}_{-19.9}  \pm0.8   ~(\textless 231.1 )  &3  .2^{+9.1 }_{-1.6}  \pm0.1   ~(\textless 11.0)  &~0.7\\
        4.781 & 511.5  & 1.06 & 0.25 &3.5  _{-2.3 }^{+4.0}   ~(\textless 11.1) &99   .4^{+113.6}_{-65.3}  \pm4.0   ~(\textless 315.3 )  &7  .3^{+4.2 }_{-2.4}  \pm0.1   ~(\textless 13.0)  &~1.5\\
        4.918 & 207.8  & 1.06 & 0.25 &0.5  _{-0.5 }^{+3.1}   ~(\textless 6.4 ) &35   .2^{+218.0}_{-35.2}  \pm1.4   ~(\textless 450.0 )  &4  .5^{+14.0}_{-2.3}  \pm0.1   ~(\textless 16.2)  &~0.7\\
        4.951 & 159.3  & 1.06 & 0.25 &3.0  _{-1.9 }^{+2.6}   ~(\textless 8.0 ) &274  .3^{+237.7}_{-173.7} \pm11.0  ~(\textless 731.5 )  &12 .7^{+5.5 }_{-4.0}  \pm0.3   ~(\textless 20.8)  &~1.9\\ 
\hline
\hline
\end{tabular}}
\label{tab:signal:yields:DD}
\end{center}
\end{table}

\section{Systematic uncertainty}
The systematic uncertainties on the Born cross section measurements
mainly originate from the integrated luminosity,
$\Sigma^+(\bar{\Sigma}^-)$ reconstruction, background, angular
distribution, branching fractions, and input line shape.

\subsection{Luminosity}
The luminosity at all energy points is measured using Bhabha
events with the uncertainties of 1.0\% \cite{BESIII:2015qfd} below
4.0 GeV, 0.7\%~\cite{BESIII:2022dxl} from 4.0 to 4.6 GeV and
0.5\%~\cite{BESIII:2022ulv} above 4.6 GeV, which are taken as the
systematic uncertainties due to the luminosity measurement.

\subsection{$\Sigma^+(\bar{\Sigma}^-)$ reconstruction}
The systematic uncertainty due to the $\Sigma^+(\bar{\Sigma}^-)$
reconstruction efficiency incorporating the tracking efficiencies,
$\pi^0$ reconstruction, and $\Sigma^+(\bar{\Sigma}^-)$ mass windows,
is estimated by the control sample of $\psi(3686)\to\Sigma^+\bar\Sigma^-$ with the
same method as described in refs.~\cite{BESIII:2016nix, BESIII:2016ssr, BESIII:2019dve, BESIII:2020ktn, BESIII:2021aer, BESIII:2021gca, BESIII:2022mfx, BESIII:2022lsz, BESIII:2023lkg, BESIII:2022kzc}. The efficiency difference correlated with the angular distribution ($\cos\theta$) between data and MC is taken as the systematic uncertainty.

\subsection{Background} 
The systematic uncertainty associated with the background, which is
estimated based on the sidebands, is determined by shifting the
sideband region inward or outward by $1\sigma$. Since the number of
events for each energy point is limited, all energy points are
combined in the estimation. The maximum difference before and after
moving the sideband region is taken as the systematic uncertainty.

\subsection{Angular distribution} 
Since there are not enough events to determine the angular
distribution for each energy point separately, a control sample with
large statistics and a DIY model~\cite{evtgen2} with customizable
angular distribution are used. With a large sample of $\psi(3770)$
events as the control sample, the angular distribution is
obtained by a maximum likelihood fit to the helicity
amplitude~\cite{Faldt:2017kgy, BESIII:2021cvv, BESIII:2023euh}. The DIY sample is initially generated
with the central value of the fitting, and its efficiency is
considered as the nominal result. Subsequently, two additional DIY
samples are generated using the upper and lower limits of the fit
uncertainty. The maximum difference in efficiency between them and the
nominal one is taken as the systematic uncertainty.

\subsection{Branching fractions} 
The uncertainty from the branching fraction of $\Sigma^+\to p\pi^0$ is
0.58\% from the PDG~\cite{PDG2020}, and the uncertainty of the
branching fraction of $\pi^0\to\gamma\gamma$ is 0.03\%. Combining with
the branching fraction of the opposite side $\bar{\Sigma}^-$ decay,
the systematic uncertainty from the branching fractions is 0.8\%.

\subsection{Input line shape}
The systematic uncertainty of the input line shape is estimated by
varying the central value of the nominal input line shape within
$\pm1\sigma$ of the statistical uncertainty, and the
$\epsilon\cdot(1+\delta)$ value for each energy point is
recalculated. This process is repeated 200 times, after which a
Gaussian function is used to fit the $\epsilon\cdot(1+\delta)$
distribution. The width of the Gaussian function is taken as the
corresponding systematic uncertainty.

\subsection{Total systematic uncertainty}
The various systematic uncertainties on the Born cross section
measurement are summarized in table~\ref{systematic}. Assuming all
sources are independent, the total systematic uncertainty on the cross
section measurement is determined by adding them in
quadrature.

\begin{table}[!hpt]
	\begin{center}
	\caption{Systematic uncertainties (in \%) and their sources for each energy point on the Born cross section measurement. Here, AD denotes angular distribution, ${\cal{B}}$ denotes branching fraction, and ILS denotes input line shape.}
        \label{systematic}
	\resizebox{0.95\columnwidth}{!}{
        \begin{tabular}{ccccccccc} \hline\hline
	$\sqrt{s}$ (GeV) &Luminosity &$\Sigma^+(\bar{\Sigma}^-)$ reconstruction &Background  &AD &${\cal{B}}$  &ILS &Total     \\ 
        \hline 
       From 3.510 to 3.896 & 1.0 & 1.9 & 2.7 & 2.1 & 0.8 & 0.1 & 4.1   \\
       From 4.008 to 4.600 & 0.7 & 1.9 & 2.7 & 2.1 & 0.8 & 0.1 & 4.1   \\ 
       From 4.628 to 4.951 & 0.5 & 1.9 & 2.7 & 2.1 & 0.8 & 0.1 & 4.0   \\
	\hline
        \hline
	\end{tabular}}
	\end{center}
\end{table}

\section{Fit to the dressed cross section}
The potential resonances in the line shape of the cross section for
the  $e^+e^-\to\Sigma^+\bar\Sigma^-$ reaction are searched for by fitting the dressed
cross section, $\sigma^{\rm dressed} =\sigma^{B}/|1-\Pi|^2$ (including
the VP effect), using the least $\chi^{2}$ method:
\begin{equation}
        \chi^{2} = \Delta X^{T}V^{-1}\Delta X.
\end{equation}
This is done considering the covariance matrix $V$ and the vector
of residuals $\Delta X$ between the measured and fitted cross
sections. The covariance matrix incorporates both the correlated and
uncorrelated uncertainties among different energy points. The
systematic uncertainties associated with the luminosity,
$\Sigma^+(\bar{\Sigma}^-)$ reconstruction, and branching fraction are
assumed to be fully correlated among the CM energies, while the other
systematic uncertainties are assumed to be uncorrelated.

\begin{figure}[!hbpt]
	\begin{center}
        \includegraphics[width=0.48\textwidth]{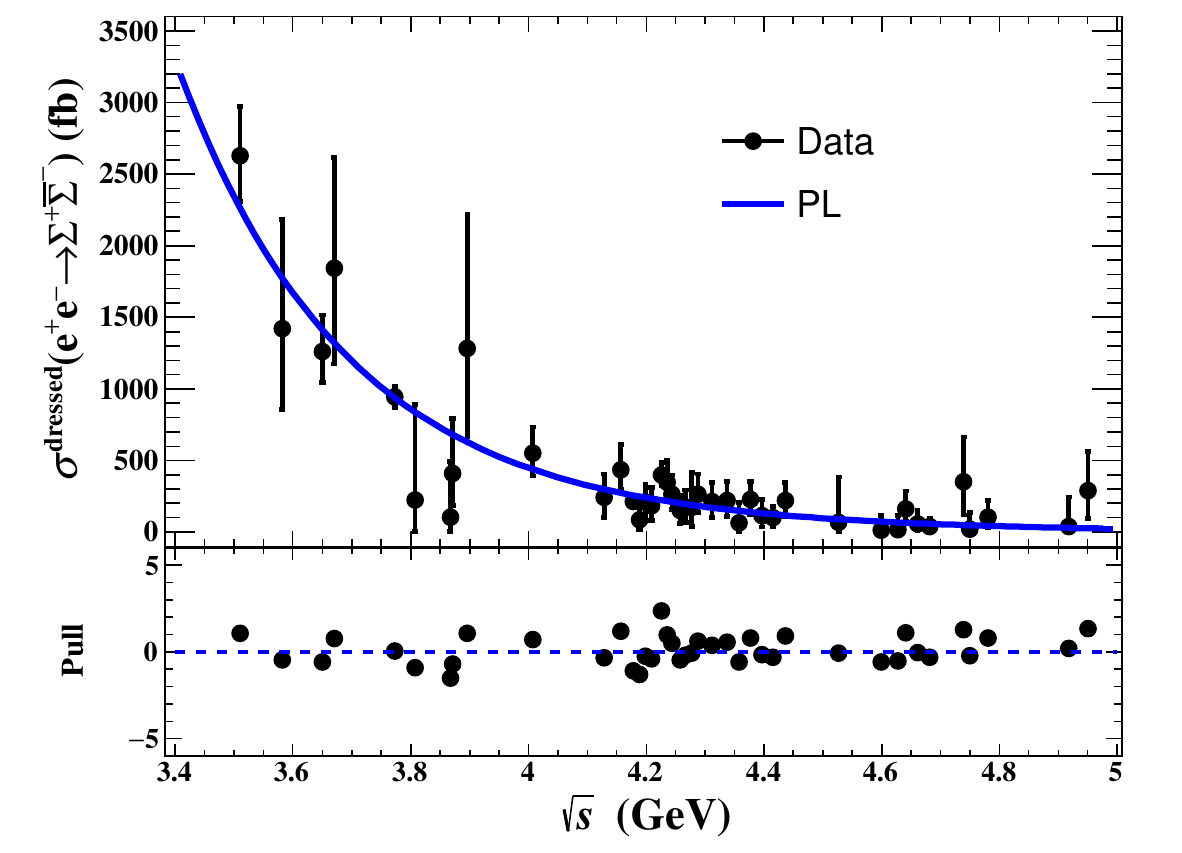}
        \includegraphics[width=0.48\textwidth]{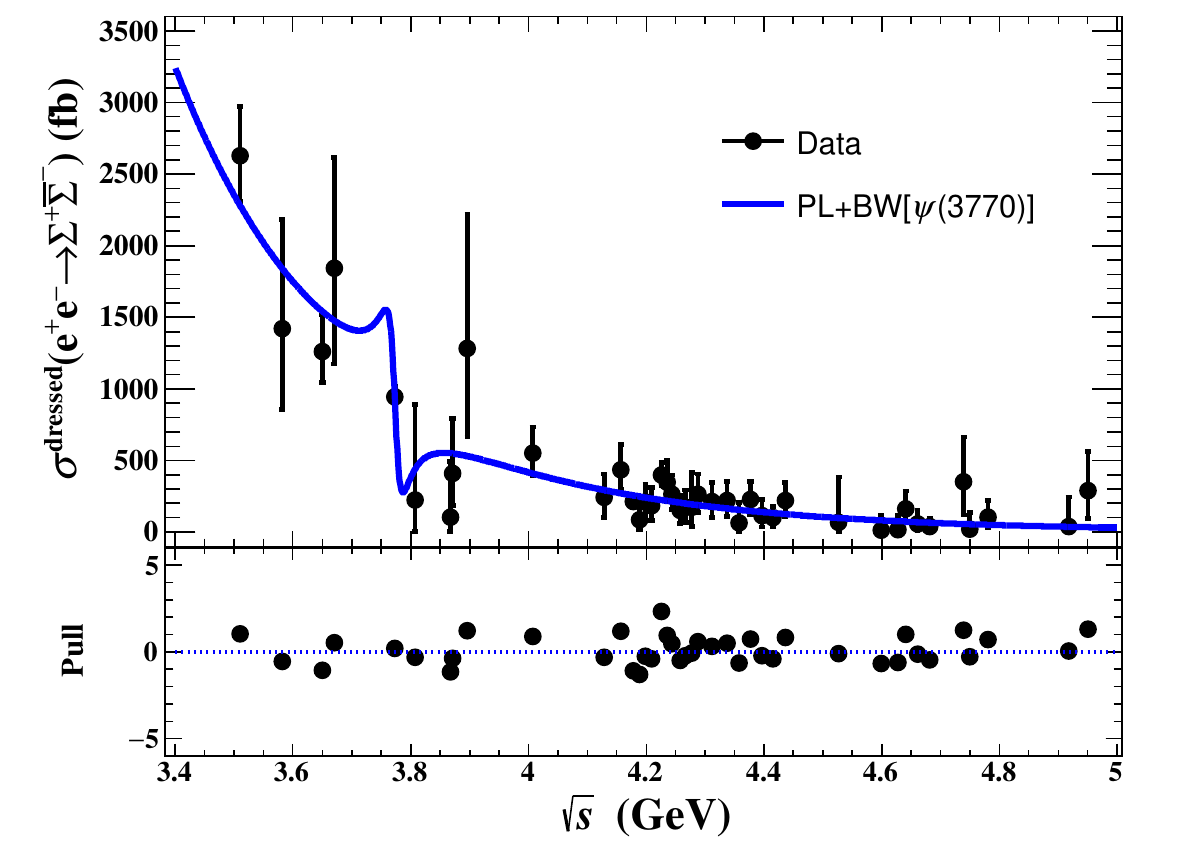}\\
        \includegraphics[width=0.48\textwidth]{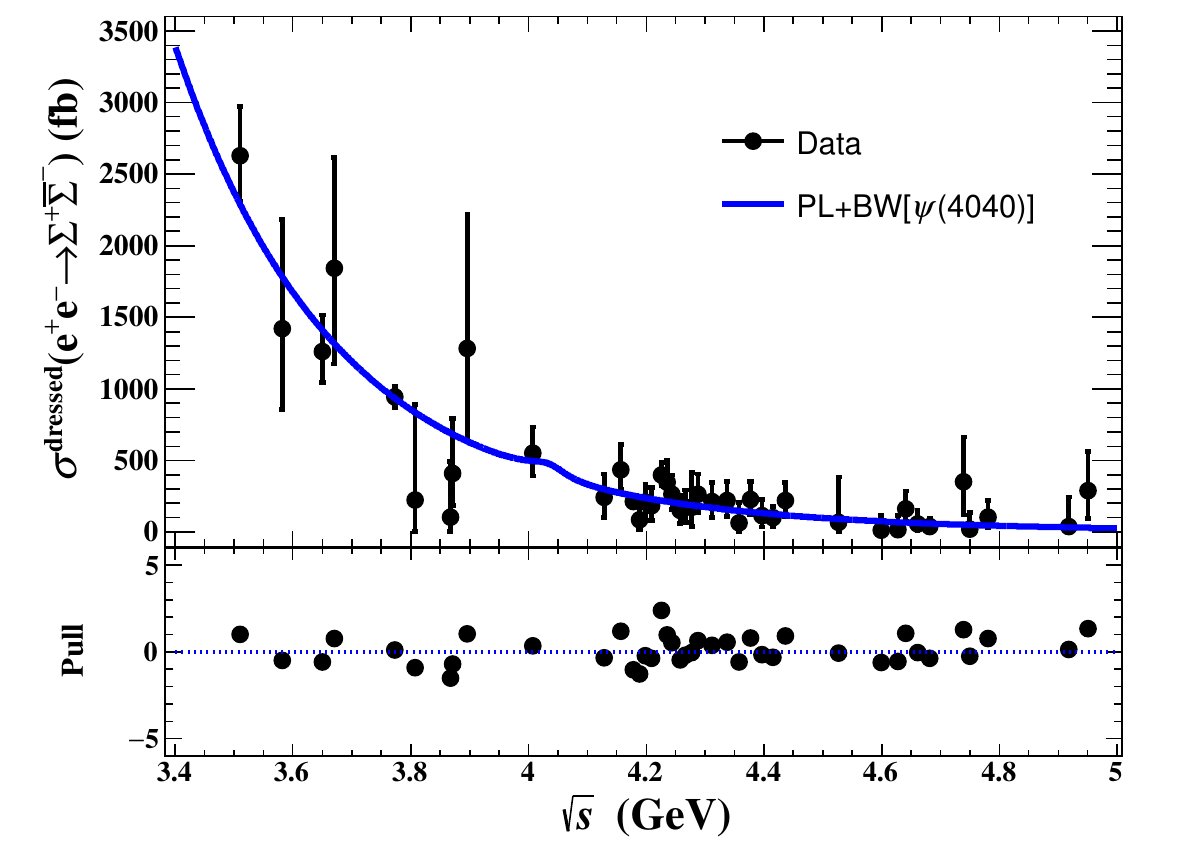}
        \includegraphics[width=0.48\textwidth]{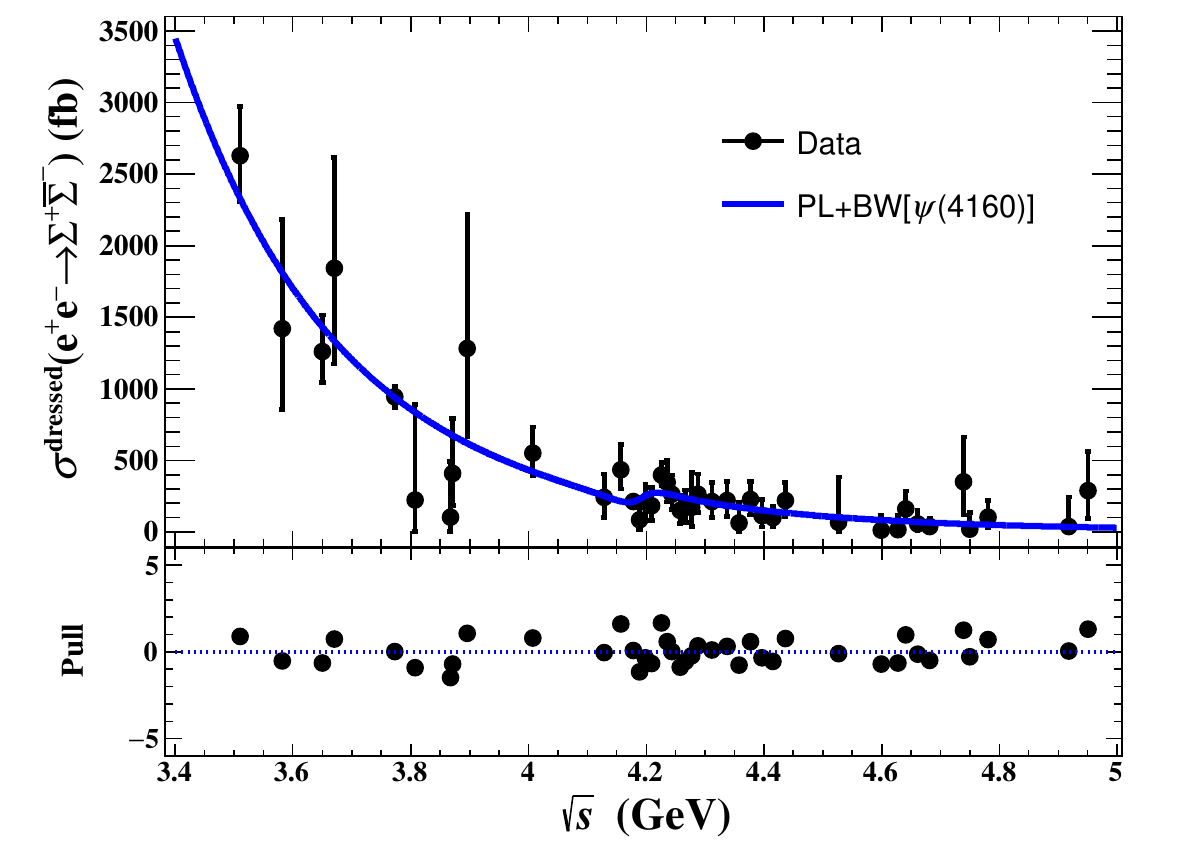}\\
        \includegraphics[width=0.48\textwidth]{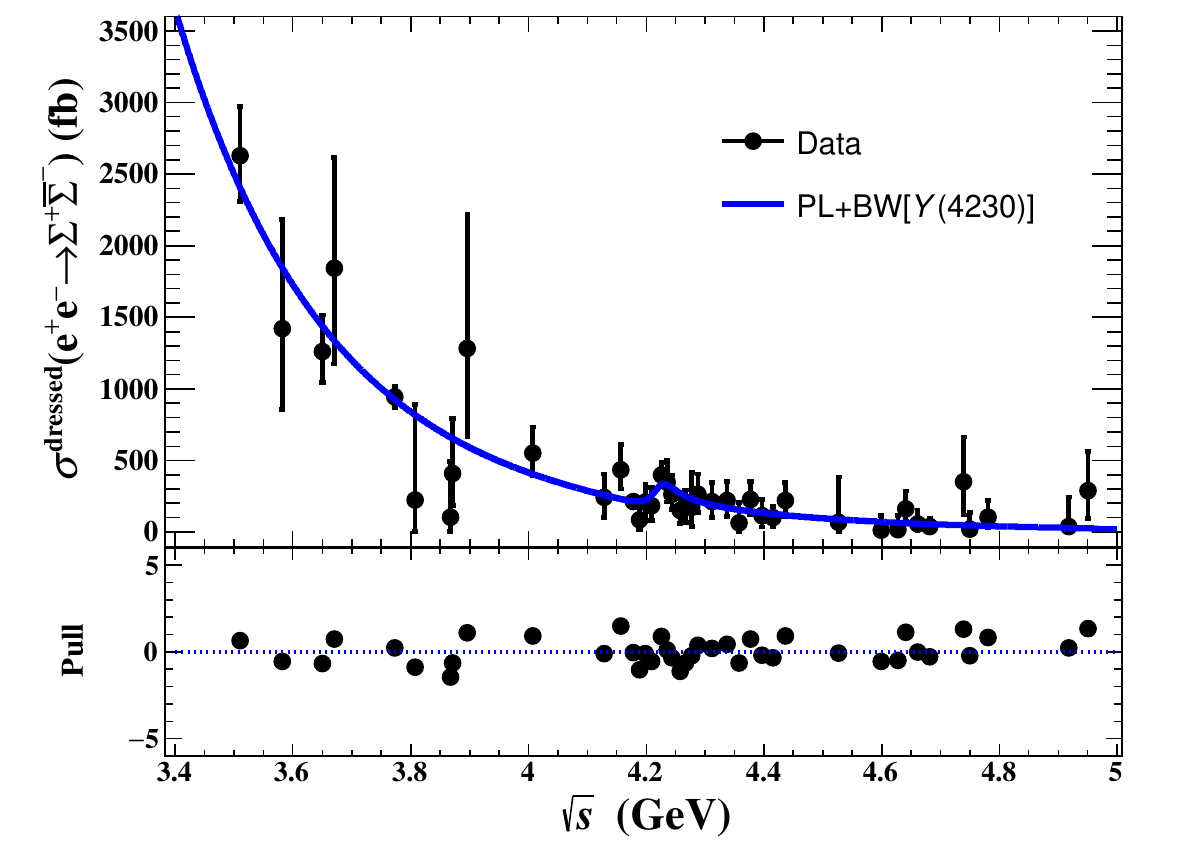}
        \includegraphics[width=0.48\textwidth]{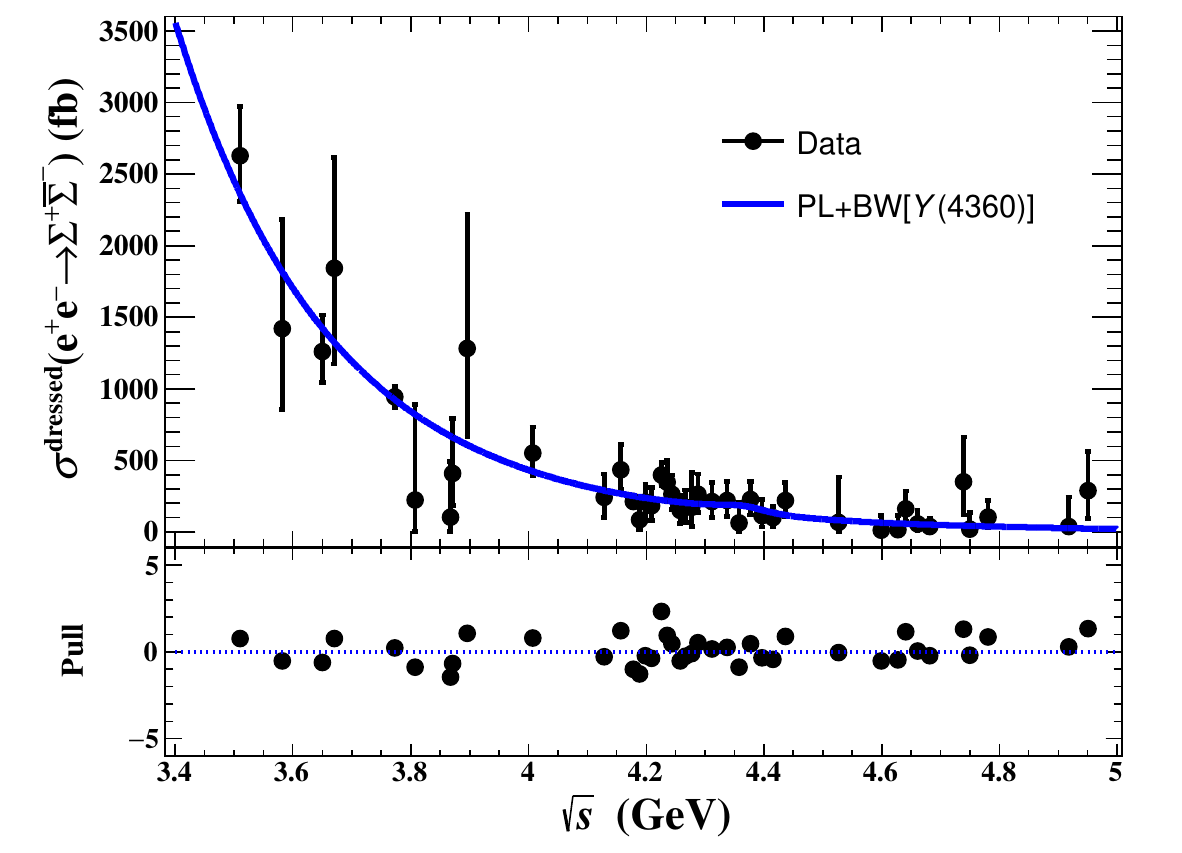}\\
        \includegraphics[width=0.48\textwidth]{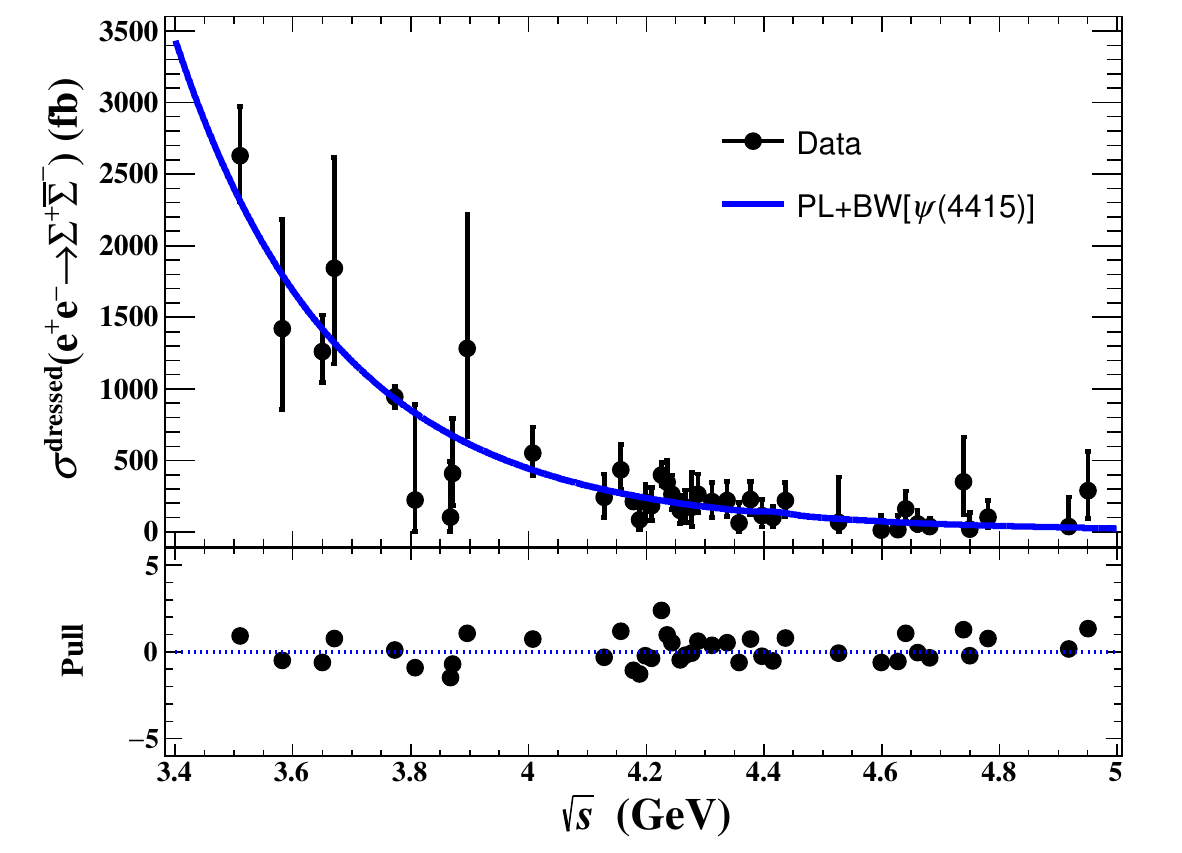}
        \includegraphics[width=0.48\textwidth]{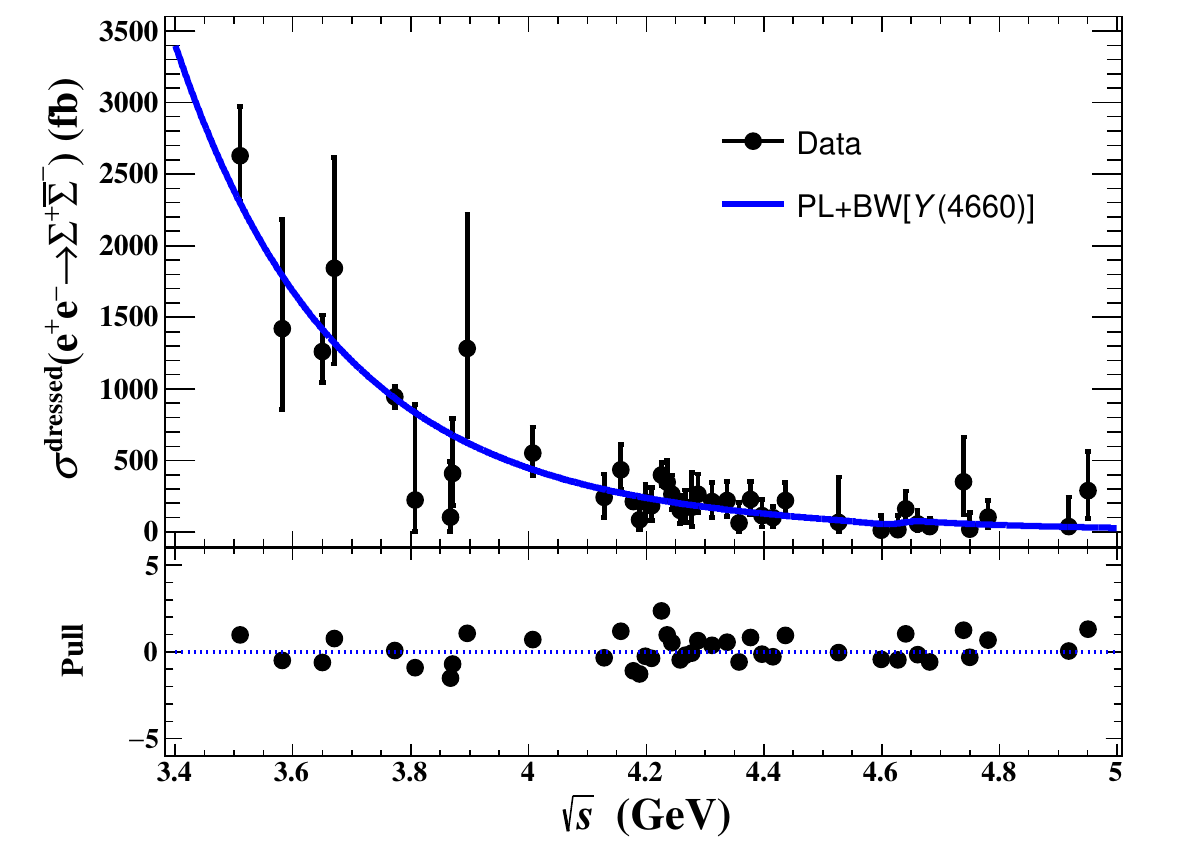}\\
	\end{center}
	\caption{Fits to the dressed cross section at the CM energy
          from 3.510 to \SI{4.951}{GeV} with the assumptions of a PL
          function only (upper left) and a power-law function plus a resonance
          ($\psi(3770)$, $\psi(4040)$, $\psi(4160)$, $Y(4230)$,
          $Y(4360)$, $\psi(4415)$, and $Y(4660)$). Dots with error
          bars are the dressed cross sections, and the solid lines show
          the fit results. The error bars represent the statistical
          and systematic uncertainties summed in quadrature.}
	\label{Fig:XiXi::CS::Line-shape-3773}
\end {figure}

Assuming the cross section of  $e^+e^-\to\Sigma^+\bar\Sigma^-$ includes a resonance plus a
continuum contribution, a fit to the dressed cross section with the
coherent sum of a power-law (PL) function plus a Breit-Wigner (BW)
function
 \begin{equation}\label{BCS_1}
	\sigma^{\rm dressed}(\sqrt{s})= \left|{\rm PL}(\sqrt{s}) + e^{i\phi}{\rm BW}(\sqrt{s})\sqrt{\frac{P(\sqrt{s})}{P(M)}}\right|^{2},
 \end{equation}
is applied. Here $\phi$ is the relative phase between the BW function
  \begin{equation}
{\rm BW}(\sqrt{s}) =\frac{\sqrt{12\pi\Gamma_{ee}{\cal{B}}\Gamma}}{s-M^{2}+iM\Gamma},
 \end{equation} and the PL function
 \begin{equation}
{\rm PL}(\sqrt{s})=\frac{c_0\sqrt{P(\sqrt{s})}}{\sqrt{s}^n},
 \end{equation} where $c_0$ and $n$ are free fit
 parameters, $\sqrt{P(\sqrt{s})}$ is the two-body PHSP factor, the
 mass $M$ and total width $\Gamma$ are fixed to the assumed resonance
 with the PDG values~\cite{PDG2020}, and $\Gamma_{ee}{\cal{B}}$ is the
 products of the electronic partial width and the branching fraction
 for the resonance decaying into the  $\Sigma^+\bar\Sigma^-$ final state. Note 
 that due to limited statistics, we only take into account the interference between 
 the continuum contribution and each resonance, and no longer consider the interference 
 between resonances. The
 parameters without a resonance are fitted to be ($c_0 = 2.3 \pm 0.8,
 n = 8.5 \pm 0.3$) with the goodness-of-fit $\chi^{2}/n.d.f=31.2/(41-2)$,
 and the parameters including a resonance are summarized in
 table~\ref{tab:multisolution}. Considering systematic uncertainties,
 the significance for each resonance is calculated by comparing the
 change of $\chi^{2}/n.d.f$ with and without the
 resonance. Charmonium(-like) states, $\psi(3770)$, $\psi(4040)$,
 $\psi(4160)$, $Y(4230)$, $Y(4360)$, $\psi(4415)$, $Y(4660)$, are all
 fitted separately by eq.~(\ref{BCS_1}), but no significant resonance
 is found. Thus, upper limits of the products of branching fraction
 and two-electronic partial width for these charmonium(-like) states
 decaying into the  $\Sigma^+\bar\Sigma^-$ final state including the systematic
 uncertainty are determined at the 90\% C.L. using a Bayesian approach
 \cite{Zhu:2008ca}. Figure~\ref{Fig:XiXi::CS::Line-shape-3773} shows
 the fit to the dressed cross section including a resonance
 [{\it i.e.} $\psi(3770)$, $\psi(4040)$, $\psi(4160)$, $Y(4230)$, $Y(4360)$,
  $\psi(4415)$ and $Y(4660)$] and without a resonance. Due to the quadratic form of the cross section like 
eq.~(\ref{BCS_1}), there are multiple solutions~\cite{Bai:2019jrb}, which can be 
determined by scanning the parameters $\phi$ and $\Gamma_{ee}\mathcal{B}$, similar to 
the method used in ref.~\cite{BESIII:2021ccp}. The fit results and their multiple 
solutions are summarized in table~\ref{tab:multisolution}.

\begin{table}[htbp]
    \small
	\caption{The fitted resonance parameters to the dressed cross
          section for the  $e^+e^-\to\Sigma^+\bar\Sigma^-$ process with two solutions. The
          fit procedure includes both statistical and systematic
          uncertainties except for the CM energy calibration. The
          relative phase is given by $\phi$. } \centering
        \begin{tabular}{l r@{ $\pm$ }l r@{ $\pm$ }l c}
            \hline
            \hline
            Resonance parameters                                              &\multicolumn{2}{c}{Solution I}         &\multicolumn{2}{c}{Solution II}       &$\chi^2/n.d.f$\\
            \hline
            $\phi_{\psi(3770)}~(\rm{rad})$                                    &$-2.6$   &$0.4$                       &$-2.0$  &$0.4$                         &\multirow{2}{*}{$28.8/(41-4)$}\\
            $\Gamma_{ee}\mathcal{B}_{\psi(3770)} (10^{-3}\rm{eV})$            &$19.5$   &$29.1$                      &$73.8$  &$32.7~(\textless 101.5)$                                    \\
            \hline                                                                                                   
            $\phi_{\psi(4040)}~(\rm{rad})$                                    &$2.0$    &$0.6$                       &$-1.7$  &$0.1$                         &\multirow{2}{*}{$30.5/(41-4)$}\\
            $\Gamma_{ee}\mathcal{B}_{\psi(4040)} (10^{-3}\rm{eV})$            &$0.2$    &$1.3$                       &$154.6$ &$29.0~(\textless 216.6)$                                     \\
            \hline                                                                                                   
            $\phi_{\psi(4160)}~(\rm{rad})$                                    &$-0.6$   &$0.5$                       &$-1.5$  &$0.1$                         &\multirow{2}{*}{$27.9/(41-4)$}\\
            $\Gamma_{ee}\mathcal{B}_{\psi(4160)} (10^{-3}\rm{eV})$            &$0.8$    &$0.7$                       &$82.1$  &$5.6~(\textless 94.6)$                                      \\
            \hline                                                                                                   
            $\phi_{\psi(4230)}~(\rm{rad})$                                    &$0.4$    &$0.4$                       &$-1.5$  &$0.1$                         &\multirow{2}{*}{$26.5/(41-4)$}\\
            $\Gamma_{ee}\mathcal{B}_{\psi(4230)} (10^{-3}\rm{eV})$            &$1.2$    &$0.8$                       &$60.1$  &$5.0~(\textless 72.4)$                                       \\
            \hline                                                                                                   
            $\phi_{\psi(4360)}~(\rm{rad})$                                    &$1.7$    &$0.7$                       &$-1.7$  &$0.1$                         &\multirow{2}{*}{$30.1/(41-4)$}\\
            $\Gamma_{ee}\mathcal{B}_{\psi(4360)} (10^{-3}\rm{eV})$            &$0.5$    &$1.0$                       &$98.4$  &$10.4~(\textless 118.8)$                                      \\
            \hline                                                                                                   
            $\phi_{\psi(4415)}~(\rm{rad})$                                    &$0.8$    &$0.6$                       &$-1.6$  &$0.1$                         &\multirow{2}{*}{$30.5/(41-4)$}\\
            $\Gamma_{ee}\mathcal{B}_{\psi(4415)} (10^{-3}\rm{eV})$            &$0.1$    &$0.5$                       &$47.7$  &$6.8~(\textless 62.1)$                                       \\
            \hline                                                                                                   
            $\phi_{\psi(4660)}~(\rm{rad})$                                    &$-0.3$   &$0.5$                       &$-1.5$  &$0.2$                         &\multirow{2}{*}{$30.6/(41-4)$}\\
            $\Gamma_{ee}\mathcal{B}_{\psi(4660)} (10^{-3}\rm{eV})$            &$0.2$    &$1.6$                       &$31.1$  &$9.3~(\textless 49.6)$                                       \\
            \hline
            \hline
        \end{tabular}
        \label{tab:multisolution}
\end{table}

\section{Summary} Using a total of \SI{24.1}{fb^{-1}} of $e^+e^-$
collision data above open-charm threshold collected with the BESIII
detector at the BEPCII collider, the process  $e^+e^-\to\Sigma^+\bar\Sigma^-$ is
studied. The Born cross sections and effective form factors are
measured at 41 CM energies that range from 3.510 to \SI{4.951}{GeV}. A
fit to the dressed cross section of the  $e^+e^-\to\Sigma^+\bar\Sigma^-$ reaction is
performed, in which the line shape is described by a series of
resonance hypotheses plus a continuum contribution, or only a
continuum contribution. However, no obvious signal of $\psi(3770)$,
$\psi(4040)$, $\psi(4160)$, $Y(4230)$, $Y(4360)$, $\psi(4415)$, or
$Y(4660)$ is found, and upper limits for the products of branching
fraction and di-electronic partial width at the 90\% C.L. for these
charmonium(-like) states decaying into the  $\Sigma^+\bar\Sigma^-$ final state are
determined. 

\newpage
\acknowledgments
The BESIII Collaboration thanks the staff of BEPCII and the IHEP computing center for their strong support. This work is supported in part by National Key R\&D Program of China under Contracts Nos. 2020YFA0406400, 2020YFA0406300; National Natural Science Foundation of China (NSFC) under Contracts Nos. 12075107, 12247101, 11635010, 11735014, 11835012, 11935015, 11935016, 11935018, 11961141012, 12025502, 12035009, 12035013, 12061131003, 12192260, 12192261, 12192262, 12192263, 12192264, 12192265, 12221005, 12225509, 12235017; 
the 111 Project under Grant No. B20063;
the Chinese Academy of Sciences (CAS) Large-Scale Scientific Facility Program; the CAS Center for Excellence in Particle Physics (CCEPP); Joint Large-Scale Scientific Facility Funds of the NSFC and CAS under Contract No. U1832207; CAS Key Research Program of Frontier Sciences under Contracts Nos. QYZDJ-SSW-SLH003, QYZDJ-SSW-SLH040; 100 Talents Program of CAS; The Institute of Nuclear and Particle Physics (INPAC) and Shanghai Key Laboratory for Particle Physics and Cosmology; European Union's Horizon 2020 research and innovation programme under Marie Sklodowska-Curie grant agreement under Contract No. 894790; German Research Foundation DFG under Contracts Nos. 455635585, Collaborative Research Center CRC 1044, FOR5327, GRK 2149; Istituto Nazionale di Fisica Nucleare, Italy; Ministry of Development of Turkey under Contract No. DPT2006K-120470; National Research Foundation of Korea under Contract No. NRF-2022R1A2C1092335; National Science and Technology fund of Mongolia; National Science Research and Innovation Fund (NSRF) via the Program Management Unit for Human Resources \& Institutional Development, Research and Innovation of Thailand under Contract No. B16F640076; Polish National Science Centre under Contract No. 2019/35/O/ST2/02907; The Swedish Research Council; U. S. Department of Energy under Contract No. DE-FG02-05ER41374


\newpage
{\bf \noindent The BESIII collaboration}\\
\\
{\small
M.~Ablikim$^{1}$, M.~N.~Achasov$^{4,c}$, P.~Adlarson$^{75}$, O.~Afedulidis$^{3}$, X.~C.~Ai$^{80}$, R.~Aliberti$^{35}$, A.~Amoroso$^{74A,74C}$, Q.~An$^{71,58,a}$, Y.~Bai$^{57}$, O.~Bakina$^{36}$
% ...
\\
\\
{\it
$^{1}$ Institute of High Energy Physics, Beijing 100049, People's Republic of China\\
% ...
$^{3}$ Bochum  Ruhr-University, D-44780 Bochum, Germany\\
$^{4}$ Budker Institute of Nuclear Physics SB RAS (BINP), Novosibirsk 630090, Russia\\
% ...
$^{24}$ Hunan Normal University, Changsha 410081, People's Republic of China\\
% ...
$^{35}$ Johannes Gutenberg University of Mainz, Johann-Joachim-Becher-Weg 45, D-55099 Mainz, Germany\\
$^{36}$ Joint Institute for Nuclear Research, 141980 Dubna, Moscow region, Russia\\
% ...
$^{57}$ Southeast University, Nanjing 211100, People's Republic of China\\
$^{58}$ State Key Laboratory of Particle Detection and Electronics, Beijing 100049, Hefei 230026, People's Republic of China\\
% ...
$^{71}$ University of Science and Technology of China, Hefei 230026, People's Republic of China\\
% ...
$^{74}$ University of Turin and INFN, (A)University of Turin, I-10125, Turin, Italy; (B)University of Eastern Piedmont, I-15121, Alessandria, Italy; (C)INFN, I-10125, Turin, Italy\\
$^{75}$ Uppsala University, Box 516, SE-75120 Uppsala, Sweden\\
% ...
$^{80}$ Zhengzhou University, Zhengzhou 450001, People's Republic of China\\
\vspace{0.2cm}
\\
$^{a}$ Deceased\\
% ...
$^{c}$ Also at the Novosibirsk State University, Novosibirsk, 630090, Russia\\
% ...
}}

\end{document}